\input harvmac
\input amssym

%DEFINITIONS
\input epsf 
 \def\IR{{ \Bbb R} }

 \def\Tr{{\rm Tr}} 
\def\hf{ {\textstyle{1\over 2}} }
 \def\frac#1#2{{\textstyle{#1\over#2}}} 
\def\({\left(} 
\def\){\right)} 
\def\<{\left\langle\,} 
\def\>{\, \right\rangle} 
\def\[{\left[} 
\def\]{\right]} 
\def\p{\partial}

\def\d{\delta} 
\def\e{\epsilon}

\def\l{\lambda} 
\def\m{\mu} 
 
\def\x{\xi} 
\def\r{\rho} 
\def\s{\sigma} 
\def\t{\tau} 
\def\z{\zeta }

\def\tL{ \tilde \L }
 
\def\L{\Lambda} 
\def\O{\Omega} 
  \def\CS {{\cal S}} 
\def\Det{{\rm Det }} 
\def\Xp{{X}_{_{+}}} 
\def\Xm{{X}_{_{-}}} 
\def\Xpm{{ X}_{_{\pm}}} 

\def\eh{\hat{\epsilon}} 
\def\xp{x_{_{+}}} 
\def\xm{x_{_{-}}} 
\def\xpm{x_{_{\pm}}} 
 
\def\xip{\xi_{_{+}}} 
\def\xim{\xi_{_{-}}} 
\def\xipm{\xi_{_{\pm}}}

 %REFERENCES
 
\lref\reviews{ I. Klebanov, {\it String theory in two dimensions},
 [hep-th/9108019; P. Ginsparg and G. Moore, {\it Lectures on 2D
gravity and 2D string theory},  [hep-th/9304011; J.~Polchinski,
{\it What is string theory?}, [hep-th/9411028].  }
\lref\KM{ V.~A.~Kazakov and A.~A.~Migdal, {\it Recent Progress In The
Theory Of Noncritical Strings,} Nucl.\ Phys.\ B {\bf 311}, 171
(1988).  }
\lref\KKK{ V.~Kazakov, I.~K.~Kostov and D.~Kutasov, {\it A matrix model
for the two-dimensional black hole,} Nucl.\ Phys.\ B {\bf 622} (2002)
141 [ [hep-th/0101011].  }
\lref\twodbh{ S.~Elitzur, A.~Forge and E.~Rabinovici, {\it Some global
aspects of string compactifications,} Nucl.\ Phys.\ B {\bf 359}, 581
(1991).  K.~Bardacki, M.~J.~Crescimanno and E.~Rabinovici,
{\it Parafermions From Coset Models} Nucl.\ Phys.\ B {\bf 344}, 344
(1990).  M.~Rocek, K.~Schoutens and A.~Sevrin, {\it Off-shell WZW models
in extended superspace,} Phys.\ Lett.\ B {\bf 265}, 303 (1991).
G.~Mandal, A.~M.~Sengupta and S.~R.~Wadia, {\it Classical solutions of
two-dimensional string theory,} Mod.\ Phys.\ Lett.\ A {\bf 6}, 1685
(1991).  E.~Witten, {\it On string theory and black holes,} Phys.\ Rev.\
D {\bf 44}, 314 (1991).  }
\lref\GKt{ D.~J.~Gross and I.~R.~Klebanov, {\it One-Dimensional String
Theory On A Circle,} Nucl.\ Phys.\ B {\bf 344}, 475 (1990).  }
\lref\GKv{ D.~J.~Gross and I.~R.~Klebanov, {\it Vortices And The
Nonsinglet Sector Of The C = 1 Matrix Model,} Nucl.\ Phys.\ B {\bf
354}, 459 (1991).  }
\lref\BULKA{ D.~Boulatov and V.~Kazakov, {\it One-dimensional string
theory with vortices as the upside down matrix oscillator,} Int.\ J.\
Mod.\ Phys.\ A {\bf 8} (1993) 809  [hep-th/0012228].  }
\lref\MooreSG{
  G.~W.~Moore,
  {\it Gravitational phase transitions and the Sine-Gordon model,}
   [hep-th/9203061].
}
\lref\streq{
  I.~K.~Kostov,
  {\it String equation for string theory on a circle,}
  Nucl.\ Phys.\ B {\bf 624}, 146 (2002)
  [hep-th/0107247].
 }
\lref\Malong{ J.~Maldacena, {\it Long strings in two dimensional string
theory and non-singlets in the matrix model,} JHEP {\bf 0509} (2005)
078 [hep-th/0503112].  }
\lref\FZZb{V.~Fateev, A.~B.~Zamolodchikov and A.~B.~Zamolodchikov,
{\it Boundary Liouville field theory.  I: Boundary state and boundary
two-point function,}  hep-th/0001012].  }
\lref\Fid{L.~Fidkowski, {\it Solving the eigenvalue problem arising from
the adjoint sector of the c = 1 matrix model,} hep-th/0506132].
}
\lref\MP{ Some references where similar system was studied are,
J.~A.~Minahan and A.~P.~Polychronakos, {\it Integrable systems for
particles with internal degrees of freedom,} Phys.\ Lett.\ B {\bf
302}, 265 (1993) [hep-th/9206046]; Phys.\ Lett.\ B {\bf 326},
288 (1994) [hep-th/9309044]; I.~Aniceto and A.~Jevicki, {\it Notes
on Collective Field Theory of Matrix and Spin Calogero Models,}
hep-th/0607152].}
\lref\Gaiotto{ D.~Gaiotto, {\it Long strings condensation and FZZT
branes,} [hep-th/0503215].  }
 \lref\AKK{ S.~Y.~Alexandrov, V.~A.~Kazakov and I.~K.~Kostov,
 {\it Time-dependent backgrounds of 2D string theory,} Nucl.\ Phys.\ B
 {\bf 640}, 119 (2002) [hep-th/0205079].  }
 \lref\flows{I.~K.~Kostov, {\it Integrable flows in c = 1 string
 theory,} J.\ Phys.\ A {\bf 36}, 3153 (2003) [Annales Henri Poincare
 {\bf 4}, S825 (2003)] [hep-th/0208034].  }
\lref\NMM{
  S.~Y.~Alexandrov, V.~A.~Kazakov and I.~K.~Kostov,
  {\it 2D string theory as normal matrix model,}
  Nucl.\ Phys.\ B {\bf 667}, 90 (2003)
  [hep-th/0302106].
  }
 \lref\Sergth{ S. Alexandrov, {\it Matrix Quantum Mechanics and
Two-dimensional String Theory in Non-trivial Backgrounds,}
 PhD Thesis, 
[hep-th/0311273].  }
 \lref\Yin{
  X.~Yin,
   {\it Matrix models, integrable structures, and T-duality of type 0 string
  theory,}
  Nucl.\ Phys.\ B {\bf 714}, 137 (2005)
  [hep-th/0312236].
}
\lref\AlexandrovCG{ S.~Y.~Alexandrov and I.~K.~Kostov,
{\it Time-dependent backgrounds of 2D string theory: Non-perturbative
effects,} JHEP {\bf 0502}, 023 (2005) [hep-th/0412223].  }

\lref\ADKMV{ M.~Aganagic, R.~Dijkgraaf, A.~Klemm, M.~Marino and
C.~Vafa, {\it Topological strings and integrable hierarchies,} PhD
thesis, [hep-th/0312085].  }
\lref\Tesc{ J.~Teschner, {\it On Tachyon condensation and open-closed
duality in the c = 1 string theory,} JHEP {\bf 0601}, 122 (2006)
[hep-th/0504043].  }
\lref\MS{
  J.~M.~Maldacena and N.~Seiberg,
  {\it Flux-vacua in two dimensional string theory,}
  JHEP {\bf 0509}, 077 (2005)
  [hep-th/0506141].
}
\lref\Seib{
  N.~Seiberg,
   {\it Long strings, anomaly cancellation, phase transitions, T-duality and
  locality in the 2d heterotic string,}
  JHEP {\bf 0601}, 057 (2006)
  [hep-th/0511220].
}
\lref\Mukh{
  A.~Mukherjee and S.~Mukhi,
   {\it Noncritical string correlators, finite-N matrix models and the vortex
  condensate,}
  JHEP {\bf 0607}, 017 (2006)
  [hep-th/0602119].
 }
\lref\DWJ{
  J.~L.~Davis, F.~Larsen, R.~O'Connell and D.~Vaman,
  {\it Integrable deformations of c-hat = 1 strings in flux backgrounds,}
  arXiv:hep-th/0607008].
  }
 \lref\HCIZ{ Harish Chandra, A.J.Math {\bf 80} 241 (1958); C.Itzykson
 and J.B.Zuber, J.Math.Phys.  {\bf 21} 411 (1980) }
\lref\Shatash{ S.~L.~Shatashvili, {\it Correlation functions in the
Itzykson-Zuber model,} Commun.\ Math.\ Phys.\ {\bf 154}, 421 (1993)
[hep-th/9209083].  }
\lref\moroz{ A. Morozov, ÒPair correlator in the Itzykson-Zuber
IntegralÓ, Modern Phys.  Lett.  A 7, no.  37 3503Ð3507 (1992).  }
\lref\BE{ M.~Bertola and B.~Eynard, {\it Mixed correlation functions of
the two-matrix model,} J.\ Phys.\ A {\bf 36}, 7733 (2003)
[hep-th/0303161].  }
\lref\EynM{ B.~Eynard, {\it A short note about Morozov's formula,}
[math-ph/0406063].  }
\lref\EF{B.~Eynard and A.~P.~Ferrer, {\it 2-matrix versus complex matrix
model, integrals over the unitary group as triangular integrals,}
[hep-th/0502041].  }
\lref\EO{ B. Eynard, N. Orantin, {\it Mixed correlation functions in the
2-matrix model, and the Bethe Ansatz,} [hep-th/0504029].  }
 \lref\HM{ Y.~Hatsuda and Y.~Matsuo, {\it Symmetry and integrability of
 non-singlet sectors in matrix quantum mechanics,}
 [hep-th/0607052].  }
\lref\Jevicki{ A.~Jevicki, {\it Development in 2-d string theory,}
[hep-th/9309115].  }
\lref\FZZ{V. Fateev, 
A. Zamolodchikov and Al. Zamolodchikov, {\it unpublished}.}
 
  \lref\FKV{ L. Faddeev, R. Kashaev and A. Volkov, {\it Strongly coupled quantum discrete Liouville theory. I: 
Algebraic approach and duality}, Commun. Math. Phys. 219 (2001) 199 [hep-th/0006156]. 
}
 
 %%%%%%%%%%%%%%%%%%%%%% 
\overfullrule=0pt 
\Title{\vbox{\baselineskip12pt\hbox 
{SPhT-T06/116}\hbox{hep-th/0610084} }} 
{\vbox{\centerline{Long Strings and Chiral Non-Singlets } 
\vskip2pt 
\centerline{ in Matrix Quantum Mechanics   } 
}} 

 \centerline{  Ivan Kostov  }

 \vskip 0.7 cm 
 
  \centerline{ \vbox{\baselineskip12pt\hbox 
 {\it Service de Physique Th{\'e}orique, CNRS -- URA 2306,}
  \hbox{ {\it \ \ \ C.E.A. - Saclay, F-91191 Gif-Sur-Yvette, France }
  }}}

%%%%%%%%%%%%%% 
 
\vskip 1cm 
 
 \baselineskip=11pt 
{\ninepoint 
\noindent{ 
We study the non-singlet sectors of Matrix Quantum Mechanics in
application to two-dimensional string theory.  We use the chiral
formalism, which operates directly with incoming and outgoing
asymptotic states, related by a scattering operator.  We argue that
a general non-singlet scattering amplitude decomposes into scattering 
amplitudes in the adjoint. In the adjoint  representation, we express the 
phase of the  tree-level scattering amplitude as an integral over  the Fermi 
sea.  We claim that this expression holds for any, in general  time-dependent, 
profile of the Fermi sea. In the case of stationary Fermi sea our formula 
reproduces Maldacena's scattering amplitude for a long string to go in and 
come back to infinity.
}}

%%\draft 
\Date{} 
%\vfill 
%\eject 
 
%%%%%%%%%%%%%%%%%%%%%%%%%%%%%% 
% 
\baselineskip=16pt plus 1pt minus 2pt

\newsec{Introduction}

\noindent 
The two-dimensional string theory arises as the collective field
theory of the gauged matrix quantum mechanics in ``upside-down''
gaussian potential, or shortly MQM \reviews.\foot{The relation between
the 2D string theory and MQM was originally proposed by Kazakov and
Migdal \KM .} MQM involves one gauge field $A_{i}^j$ and one scalar
field $X_{i}^{j}$, both hermitian $N\times N$ matrices.  The theory is
formally described by the action
\eqn\Maction{
\CS= \int dt \, \Tr \[ P\, \nabla_A X - \hf (P^2- X^2) %+\mu
\] , }
where $\nabla_A X = \p_t X -i [A, X]$ is the covariant time
derivative.  Since the potential is bottomless,
one should  introduce a cutoff and tune the size $N$ with the cutoff 
before taking the large $N$ limit.

The tachyon dynamics in 2D string theory with flat, linear dilaton,
background is described by the {\it singlet sector} of the Hilbert space.
In the singlet sector the gauge field simply plays the role of
Lagrange multiplier assuring that the matrix coordinate and momentum
can be simultaneously diagonalized.  As a consequence, the action
\Maction\ describes a system of $N$ non-relativistic free fermions in
the upside-down quadratic potential.  The ground state of the system
is characterized by the Fermi level $E_F=-\mu$, where $g_s = 1/\mu$ is
the string coupling  constant.

The states obtained by introducing one or more Wilson lines in the
adjoint representation belong to the {\it non-singlet sector} of the
Hilbert space.  The wave functions in this sector transform by
non-trivial representations of $SU(N)$ that can be obtained by
tensoring the adjoint.  Besides tachyons the non-singlet sector
contains excitations which in the compactified Euclidean theory
correspond to winding modes, or vortices, on the world sheet
\refs{\GKt, \GKv, \BULKA}.  The non-singlet sector may contain more
general string backgrounds.  It was conjectured \FZZ\KKK\    that adding two
oppositely oriented Polyakov loops in the action \Maction,
\eqn\Poll{\d \CS =\l \Tr \exp \(\int _{0}^R dt A(t)\) + \l \Tr \exp
\(-\int _{0}^R dt A(t)\) , }
may deform the metric of the target space and produce Euclidean black
hole background \twodbh.  It is known that the deformations  by   winding  modes are 
integrable \MooreSG\ and  the deformed partition 
function is a  tau-function of a non-compact Toda hierarchy \KKK\streq .
However,  knowing  the partition function is not sufficient to reproduce the 
geometry of the background.  In  order to probe the background curvature,
one should be able to calculate scattering amplitudes of tachyons in presence of a 
deformation by a non-singlet source. This problem is well posed in the Lorentzian
formulation of the theory.
  
Recently, Maldacena \Malong\ gave a world-sheet interpretation of the
non-singlets in Lorentzian MQM. He considered mainly the states
containing one Wilson line, which transform according to the simplest
non-trivial representation, the adjoint.  According to Maldacena, the
wave function in the adjoint describes a particle-like excitation
(impurity), interacting with the Fermi sea.  The collective coordinate
of this  ``adjoint particle", describes the position of the tip of a
folded string that stretches from infinity.  This interpretation
passed several consistency tests.  The tree level scattering amplitude
for the tip of the long string, evaluated in \Malong\ using the FZZ
boundary Liouville amplitude \FZZb , turned out to be the same as the
one extracted from the wave function in the adjoint representation,
$\CR_{ij; kl}(\O) = \O_{ik} \CO^*_{jl} - \d_{ik}\d_{jl}$
subsequently calculated in \Fid.  Furthermore, the density of states
corresponding to the phase shift reproduced correctly the
vortex-antivortex correlation function in the compactified Euclidean
theory \Malong. Further study of long strings in 2D string theory is 
presented in \MS\Seib.

In order to make the next steps towards  constructing a  solvable model 
of  Lorentzian 2D black hole it is important to extend this picture  to the 
whole non-singlet sector.  As all allowed representations are contained 
in the tensor products of the adjoint,
all non-singlet states can be described in terms of tachyons and
adjoint particles, or long strings.  The wave function of such a state
can be in principle computed by solving the corresponding matrix
Calogero equation.  However, in spite of the fact that problem is
integrable \MP, this seems to be a rather hopeless task.  Some
interesting speculations concerning higher representations were
presented in \Gaiotto, but no quantitative description exists so far.

\bigskip

In this paper we argue that the scattering problem in the non-singlet
sector of MQM can be solved by means of the {\it chiral formalism}
introduced in \AKK\ and later developed in 
  \flows\AlexandrovCG\Sergth\ADKMV\Tesc\MS\Yin\DWJ\NMM\Mukh , 
  in which one performs
a canonical transformation to the light cone variables
\eqn\YpYm{ X_+ = \frac{ X+ P}{ \sqrt{2}}, \ \ \ \ X_- = \frac{X- P}{
\sqrt{2}}.  }
The wave functions in $\Xp$ and $\Xm$ spaces describe the asymptotic
states in the original theory respectively in the infinite past and
future.  The outgoing and the incoming states are related by Fourier
transformation
\eqn\Fouriertr{\eqalign{ \Phi^-(\Xm) &= \ \int d\Xp e^{ i\Tr \Xp\Xm}\,
\Phi^+(\Xp), }}
which represents the scattering operator in the chiral basis.  The
Hamiltonian corresponding to the action \Maction\ is bilinear in the
new variables,
\eqn\opham{ H=-\frac{1}{2}\Tr ( \Xp\Xm + \Xm \Xp) \, , }
and  the general solution of the corresponding  Schr\"odinger equation is  
\eqn\GENSOL{ \Phi^{\pm} (\Xpm  , t)  =   
 e^{\mp {1\over 2} N^2 t}  \ \Phi^{\pm}(e^{\mp t}\Xpm  ).}

In the non-singlet sector the wave function is a vector transforming
according to a (not necessarily irreducible) representation $\CR$ of
$SU(N)$:
\eqn\Unrot{ \Phi_I^\pm(\O \Xpm\O^\dagger)=\sum_{J=1}^{{\rm dim} \CR}
\CR_{I,J}(\O)\, \Phi^\pm_J(\Xpm), \qquad \O\in SU(N), }
where $\CR_{I,J}(\O)$ is the matrix of the representation.
The $U(N)$ symmetry allows to reduce, as in the singlet sector, the
original $N^2$ degrees of freedom to the $N$ eigenvalues $
x^\pm_1\dots x^\pm_N$ of the matrix $\Xp$ or $\Xm$.

Our claim is that one can construct a large $N$ collective field
theory for the scattering amplitude between the states $ \Phi ^-$ and
$ \Phi ^+ $, which is given by the inner product
\eqn\inner{ \( \Phi^- | \Phi^+ \)= \int d \Xp d\Xm \ e^{ i \Tr \Xp
\Xm} \ \sum_{J=1}^{{\rm dim} \CR} \ \overline{ \Phi^- _I ( \Xm)} \
\Phi ^+_I (\Xp).  }
This is achieved in two steps.  The first step consists in integrating
out the angular degrees of freedom in the matrix integration measure
\eqn\measre{ d\Xpm = d \O_\pm \, dx^\pm_1\dots dx^\pm_N\,
\Delta^2(\Xpm), }
where $ \Delta(\Xpm)= \prod _{i<j} (x^\pm_i- x^\pm_j)$ is the
Vandermonde determinant.  The second step is to formulate the result
in terms of the collective field, the phase space eigenvalue density
$\rho(x^+, x^-)$.

As a consequence of \Unrot , the  angular dependence of the integrand in 
\inner\ is only through the factor 
$$\sum_{K=1}^{{\rm dim} \CR}
 \CR_{I,K}(\O_+)\CR_{K,J}(\O_-^\dagger)=
\CR_{I,J}(\O_+\O_-^\dagger).$$
The integral   with respect to $\O=
\O_+\O_-^\dagger$  depends only on the representation $\CR$ and not on
the wave functions:
\eqn\GIZ{ I ^{\CR} _{I,J}  (\Xp, \Xm )= \int \limits_{SU(N)} \!\!\!  d \O\
\CR_{I,J} (\O) \ e^{i \Xp \O \Xm \O^{\dagger}}.}
In the singlet sector, $\CR = {\Bbb I}$, this is
Harish-Chandra-Itzykson-Zuber integral \HCIZ. The general integral
\GIZ\ has been first studied by Shatashvili \Shatash\ who found the
complete solution, but in a form not explicitly invariant under
permutations of eigenvalues and hence not immediately applicable in
the large $N$ limit.  A nice symmetric formula for the adjoint
representation,  $\CR_{ik, jl}(\O)=  \O_{ij} \O^\dagger _{lk} 
- \O_{ik}  \O^\dagger _{lj} $
was originally proposed by Morozov \moroz\ and
subsequently proved by Eynard and collaborators \BE\EynM.
Generalization of Morozov-Eynard formula for any representation was
found in \EF. Therefore the first step, the $U(N)$ integration, is
essentially accomplished.
 
In order to extract the large $N$ limit and construct the collective
field theory, we need to evaluate the inner product \inner\ for the
wave functions that describe low energy excitations above the ground
state.  The latter is constructed by to filling the fermionic energy
levels of the singlet sector up to the the Fermi level $E_F=-\mu$.  It
is sufficient to solve the problem for a standard set of functions
spanning the sector containing $p$ adjoint particles,
\eqn\wfgen{ \hat \Phi^{\pm} _{i_1\cdots i_p, j_1\cdots j_p} (
\xi^\pm_1,...,\xi^\pm _p;\Xpm) 
= \[ {1\over \x^\pm_1 + \Xpm} \]_{i_{_1}
 j_{\s^{\pm}(1)} 
} \cdots \[ {1\over \x^\pm_p + \Xpm} \]_{i_{_p}
j_{\s^\pm(p)} }\!  \Phi_0 ^\pm (\Xpm), }
where $ \Phi_0 ^\pm $ is the ground state wave function 
and $\s^\pm $ are permutations from $S_N$.  Here we suppressed
the time dependence, which can be reconstructed from \GENSOL. The
functions \wfgen\ transform according to the tensor product of $p$
adjoints.  The projection to the irreducible components of the tensor product is done by
summing over the permutations with a character of the symmetric group.

The integral for the inner product of functions of the type \wfgen\ is
similar to that for the mixed correlators in the two-matrix model,
which were evaluated in the large $N$ limit by Eynard and Orantin \EO.
The authors of \EO\ found that a generic mixed correlator decomposes
as a sum of products of two-point mixed correlators.  An important for
us fact is that the coefficients in the sum do not depend on the
radial part of the matrix measure.  Therefore the result of \EO\ must
be applicable also in the case of the inverse matrix oscillator, which
means that any scattering process involving only adjoint particles can
be decomposed into one-particle scattering amplitudes.

In this way the success of the chiral approach is guaranteed if it
works for the adjoint representation.  Our aim here is to demonstrate
that this is indeed the case.  We will obtain a formula for the
scattering amplitude of the adjoint particle in terms of an integral
over the Fermi sea, which we expect to be valid for any, in general
time-dependent, tachyon background.  In the case of stationary Fermi
sea we recover Maldacena's expression \Malong\ for the scattering
amplitude for a long string to come in and go back to infinity.  Our
more general answer can be used to calculate the tree-level processes
involving one adjoint particle and any number of tachyons.  Scattering
amplitudes involving several adjoint particles will be considered in a
elsewhere.

\newsec{The scattering amplitude in the adjoint representation}

\noindent    
It will be convenient to absorb into the wave function a Vandermonde
determinant from the measure \measre:
\eqn\PsipPhi{ \Psi^\pm (\Xpm ) : = \Delta(\Xpm ) \, \Phi ^\pm (\Xpm )
.  } 
Then a complete set of wave functions in the adjoint sector is
given by\foot{Since we do not subtract the trace, this function has a
small component in the singlet sector.  }
\eqn\Psixi{ \hat \Psi_{ij} ^\pm(\xi^\pm; \Xpm) = \[{1\over \xi ^\pm +
\Xpm} \]_{ij}\ \det_{kl} \[ \psi_{{E_k^\pm}} ^{_\pm} (x^\pm_{l})\] .
}
The last factor is the general eigenfunction of the matrix Hamiltonian
in the singlet sector, which is a Slater determinant of one-fermion
eigenfunctions
  \eqn\wavef{ \psi_{{E}} ^{_\pm}(\xpm)\ = \ \frac{1}{\sqrt{2\pi }}\
  (\xpm)^{ \pm i E-{1\over 2}} .  }
We are going to consider only states which represent incoming leftmovers and outcoming rightmovers.   Then the  eigenvalues of $\Xpm$ can be  assumed positive and the
operator functions \Psixi\ are analytic in the complex $\xi$-plane cut
along the negative real axis.  One can think of $\xi^+$ and $\xi^-$ as
the phase space coordinates of the adjoint particle.  The Hamiltonian
\opham\ acts on the wave function \Psixi\ as
\eqn\ANGH{ H \hat \Psi ^\pm = \mp i \sum_k (x^\pm_k \frac{\p}{ \p
x^\pm_k}+ \hf) \Psi ^\pm = \[ \pm i( \x^\pm \frac{\p}{\p \xi^\pm} +1)
+ E_1^\pm +\dots +E_N^\pm\] \hat \Psi ^\pm .  } We will be eventually
interested in evaluating the inner product for the eigenfunctions of
the Hamiltonian,
\eqn\adjwf{\Psi^\pm_{{\tilde E}^\pm} (\Xpm) = (\Xpm)^{\pm i {\tilde
E}^\pm }\, \det_{kl} \[ \psi_{{E_k^\pm }} ^{_\pm} (x^\pm_{l})\] , }
which are obtained as Mellin transforms of \Psixi :
\eqn\Mellin{\eqalign{ \Psi^\pm_{{\tilde E}^\pm} (\Xpm) &=
\pm{i\sinh\pi {\tilde E} \over\pi}\, \int _0^\infty d \xi \ \xi^{\pm
i{\tilde E} ^\pm } \ \Psi^\pm (\xi^\pm; \Xpm) .  } }

Our aim is to evaluate the scattering amplitude for the adjoint
particle in the large $N$ limit, when the eigenvalues can be replaced
by a collective mean field.  For that it is sufficient to consider the
diagonal elements of the scattering operator, $E^\pm_k= E_k, \
k=1,\dots, N$, which implies ${\tilde E}^+ = {\tilde E}^-$.  The
non-diagonal elements, which are of subleading order, allow to study
the back reaction of the adjoint particle to the collective field.

We will proceed as follows.  First we perform the angular integration
in the inner product of the functions \Psixi\ using Morozov-Eynard
formula.  Then we consider the large $N$ limit, in which the result
takes the form of Fredholm determinant, and evaluate the leading
contribution at large cosmological constant $\mu$.  Finally we perform
the Mellin transform according to \Mellin\ and obtain the reflection
factor as a function of the energy ${\tilde E} $.

 \subsec{Eliminating the angles}

\noindent
The angular integration in the inner product of the wave functions
\Psixi\ is evaluated by Morozov-Eynard formula \EynM
 \eqn\MorE{ \int\limits_{SU(N)} \!\!\!  d\O\, \Tr\(\!  \frac{1}{ \xim
 + \Xm} \O \frac{1}{ \xip + \Xp}\O^{\dagger}\!\) e^{\Tr (i\Xm \O \Xp
 \O^{\dagger})} = \frac{\det\(\hat S\, +i\frac{1}{\bar \xim+\Xm} \hat
 S \frac{1}{ \xip+\Xp}\) - \det \hat S }{ i\, \Delta(\Xm)\,
 \Delta(\Xp)} \, , }
where $\hat S$ is the matrix with matrix elements $ S_{jk} = e^{i
x^+_j x^-_k}$ and $\Xpm = {\rm diag}(x^\pm_1,\dots x^\pm_N)$.  
%The integral \GIZ\  for the adjoint representation 
%can be evaluated as the residue of the r.h.s. of \MorE\
%at   $\x_+ = x^+_j$ and $\x_- = x^-_k$.
 Plugging \MorE\ in the definition of the
inner product \inner, we get
\eqn\DerE{\eqalign{ (\hat \Psi^- | \hat \Psi^+ )&= -i \int_0^\infty d
^Nx^{+}d^Nx^{-} \ \overline{ \det_{kl} \[ \psi_{{E_k}} ^{_-}
(x^-_{l})\]} \det_{kl} \[ \psi_{{E_k}} ^{_+} (x^+_{l})\] \cr & \times
\det\[\hat S+i\frac{1}{\bar \xim+\Xm} \hat S \frac{1}{ \xip+\Xp}\]\, .
} }
 Here we  dropped the second  term in the
Morozov-Eynard formula, which is needed to cancel the constant term in
the expansion of the r.h.s. at $\xi_\pm\to\infty$.   
This term does not depend on 
$\xipm$ and therefore is irrelevant for the scattering phase.
 Note that the Vandermonde factors from the
measure disappear due to the redefinition \PsipPhi\ of the wave
function.

A form of the inner product suitable for taking the large $N$ limit is
obtained if we rewrite \DerE\ as a determinant of double integrals:
\eqn\DerE{\eqalign{ (\hat\Psi^- | \hat \Psi^+) &= {R^{ \rm ad}}
(\xip,\xim) \, \prod_{k=1}^N R(E_k) , } } 
\eqn\PSIAD{\eqalign{ {R^{
\rm ad}} (\xip,\xim) = -i \det _{jk} \[ \d(E_j-E_k) + i K (E_j,
E_k)\]\, , }}
 \eqn\defK{ \eqalign{ K(E', E)= & \langle E'| { 1 \over
 (\xip+\xp)(\xim+\xm)} |E\rangle \, .} }
Here $R(E)$ denotes the fermion reflection coefficient, or the bounce factor,
determined by
the inner product of the one-particle eigenfunctions \wavef,
\eqn\innerS{ {\int_0^\infty d \xp d \xm \overline{ \psi^{_{E'}} _{^-}
(x^-)} \, e^{i x_+x_-} \psi^{_{E'}} _{^+} (x^+)= R(E )\ \delta(
E-E')\, , }}
%
%Its explicit expression is
%$$
%R(E)= e^{i\phi_0(E)} =\frac{1}{\sqrt{2 \pi}}\,
%e^{-{\pi\over 2} (E- i/2)} \, \Gamma( iE + 1/2) 
%= \sqrt{1+ e^{2\pi E}} \ e^{\Im \G(i E+{1\over 2})}
%$$
%
and the r.h..s. of \defK\ is defined as
\eqn\bioO{ \< E' | f | E\>:=
{\int_0^\infty d \xp d \xm \overline{ \psi^{_{E'}} _{^-} (x^-)} \, f
(\xm,\xp) \, e^{i x_+x_-} \psi^{_{E'}} _{^-} (x^+) \over \sqrt{ R(E)\
R(E')}}.  } In particular, \eqn\biort{ \< E' |1| E\> = \delta(E'-E).
}
The second factor in the inner product \DerE\ is the inner product of
the singlet states whose wave functions are given by the Slater
determinants in \Psixi.  Therefore the first factor, $ {R^{ \rm ad}}
(\xip,\xim)$, gives the scattering amplitude (more precisely, the
reflection coefficient) for the adjoint particle as a function of the
collective coordinates, $\xip$ and $\xim$, of the initial and final
states.

\subsec{The scattering amplitude   as  a Fredholm determinant}

\noindent  The expression  \PSIAD\ for the inner product  was derived
for a generic wave function of the form   \Psixi. Now we will specify the
Slater determinant in \Psixi\ to be the wave function of the Fermi 
sea filled up to $E_F = -\mu$. For this purpose we  
introduce a large cutoff $\L\gg\mu$, say, by putting
a wall at $\xpm \sim \sqrt{\L}$, so that the spectrum becomes
discrete.  Then the density of the energy levels in the singlet sector
is given by \AKK\flows
\eqn\DEN{ \rho(E)= {\log \Lambda\over 2 \pi}- {1\over 2\pi } {d \phi_0
(E) \over d E}, \qquad \phi_0 (E) = - i \log R(E), }
 and the level spacing is \eqn\discs{ \Delta E = {2\pi\over \rho(E)}
 \approx {1\over \log\L}.  } The total number $N$ of eigenvalues is
 given by the number of energy levels between the bottom of the
 regularized Fermi sea, $E_{\rm cutoff}= -\L$, and the Fermi level
 $E_F = -\mu$:
\eqn\Nofmu{ N =\int_{-\L}^{-\m} dE\, \r(E) .  } Finally, the inner
product \innerS\ is replaced by the following relation for the
regularized eigenfunctions:
\eqn\innerD{ {\int_0^\infty d \xp d \xm \overline{ \psi^{_{E_j}} _{^-}
} \, e^{i x_+x_-} \psi^{_{E_k}} _{^+} = R(E_j ) \, \rho(E_j) \
\delta_{jk}\, .  }}

We would like to evaluate the scattering amplitude \PSIAD\ for the
case where $E_1, \dots, E_N$ are the allowed energy levels in the
interval $[-\L, - \mu]$.  In presence of a cutoff, the expression
\PSIAD\ takes the form
 \eqn\Psreg{\eqalign{ {R^{ \rm ad}} (\xip,\xim) &= -i \det _{jk} \[
 \rho(E_j) \d _{jk} + i K (E_j, E_k)\]\, \cr &=-i \prod_{k=1}^N
 \rho(E_k) \times \det _{jk} \[ \d _{jk} + i \frac{\Delta E_j}{ 2\pi}
 \ K (E_j, E_k)\]\, .  }}
The first factor is  an irrelevant  infinite constant and will be neglected.
 The second factor has a smooth limit $ \L\to \infty$:
\eqn\FrD{\eqalign{ {R^{ \rm ad}} (\xip,\xim) &= 1 + i \sum _{j}
\frac{\Delta E_j}{2\pi} \, K( E_j, E_j) -{1\over 2!} \sum _{j, k}
\frac{\Delta E_j}{2\pi} \frac{ \Delta E_k }{2\pi} \, \left| \matrix{
K( E_j, E_j) & K( E_j, E_k) \cr K( E_k, E_j) & K( E_k, E_k)}\right|
+\cdots \cr &\rightarrow 1 + i\int \limits_{-\infty}^{-\mu}\!
\frac{dE}{2\pi} \ K(E,E) -{1\over 2!} \int \limits_{-\infty}^{-\mu}\!
\frac{dE}{2\pi} \int \limits_{-\infty}^{-\mu}\!  \frac{dE'}{2\pi} \
\left| \matrix{ K( E, E) & K( E, E') \cr K( E', E) & K( E',
E')}\right|+\cdots .  } } 
The series is by definition the Fredholm determinant of the kernel
\defK, restricted to the interval $[- \infty , -\mu]$:
\eqn\Fredh{{R^{ \rm ad}} (\xip,\xim) = \Det[1+i K] .  }
 Therefore the scattering phase $S(\xip,\xim)$, defined by
\eqn\Radxi{ {R^{ \rm ad}} (\xip,\xim)=\exp\[iS(\xip,\xim)\]\, , }
 has the following integral representation: 
 \eqn\PhaseR{ S(\xip,\xim)=
 \sum _{n=1}^\infty {(-i)^{n-1}\over n} \Tr\, K ^n , }
 \eqn\defKn{\Tr \,K ^n:= \int _{-\infty}^{-\mu} \frac{dE_1}{2\pi}
 \cdots \frac{dE_n}{2\pi} \ K(E_1, E_2) \cdots K(E_n, E_1).  }

\bigskip

\subsec{Tree level calculation of the reflection factor}

\noindent
The Fredholm kernel $K(E,E')$ can be evaluated semiclassically for
large negative energies.  The easy calculation is given in Appendix A.
The result is \eqn\Kernell{\eqalign{ K(E+\e , E-\e ) &= \(\xip\over
\xim\)^{ i\e } { \(-{\xip\xim \over E}\) ^{i\e } - \(-{\xip\xim \over
E}\) ^{-i\e } \over i\sinh( 2\pi\e) }\ {\pi \over \xip\xim+E } \, .  }}
The corrections are of order $1/E$.  It is obvious that the integral
for $\Tr K^n$ is convergent for $n>1$ and behaves as $o(\m^{1-n})$.
Therefore at tree level the phase \PhaseR\ is given by the first term
of the series,
 \eqn\TrK{S(\x_+,\x_-)= \Tr K =\int_{- \tilde \L}^{-\mu} dE \ K(E,E),
 }
\eqn\sdpS{ K(E,E) =   { \ln( \xi_+ \xi_-)-\ln(-E) \over
\xi_+ \xi_- +E} .  }
 This integral is logarithmically divergent at minus infinity.  To
 make it convergent, we cut it off at $E= - \tilde \L$, where $\mu\ll
 \tilde\L\ll \L$.  The new cutoff $\tilde \L$ will be given below a
 precise meaning in terms of the world-sheet theory.  Let us introduce
 the variable
  \eqn\defs{ s =\hf \log(\x_+\x_-/\mu) }
and the function (it already appeared in Maldacena's paper)
\eqn\deff{f(x) = {1\over \pi} \int\limits_{-\infty}^{ x}d\z \({\z\over
\tanh \z } + \z\) .}
Then we can write the regularized integral \TrK\ as
 \eqn\Sexp{\eqalign{ S(\x_+,\x_-)&= f\(\hf \log(\tL /\m)- s\)- f\( -
 s\)\cr &\approx {1\over\pi} \( \hf \log(\tL /\m) - s\)^2 - f\( -s\).
 }}
In the second line we used the asymptotics $f(x)\approx x^2/\pi$ at
$x\gg1$.
 
The leading corrections to this tree-level formula come from the terms
of order $1/\mu$ in the semi-classical expression for the kernel
\sdpS, as well as from the second term, $\Tr K^2$, of the series
\PhaseR. It however natural to expect that the $1/\mu$ corrections
cancel and the perturbation series for the scattering phase is in
$1/\mu^2$.

 \subsec{ The  scattering phase as an integral over the Fermi sea}

\noindent  In the derivation of the scattering phase we did not use the 
particular form \wavef\  the one-particle functions. Therefore the derivation 
remains valid also in presence of  a tachyon source, whose only effect is 
that the  one-particle wave functions get deformed \AKK .  In this case the scattering phase 
will be given by the same expression \PhaseR, but with a deformed kernel $K$.
The  leading term, the trace \TrK, can be expressed alternatively as an
integral over the Fermi sea,
\eqn\intph{\eqalign{ S(\xip, \xim) &=\int \limits_{ 0}^{{ \infty} }
{d\xp d \xm \over 2\pi} \ {\r(\xp,\xm) \over ( \xip+\xp)( \xim +\xm) }
\ , } } 
where $\r(\xp,\xm) $ is the semi-classical density of the
fermionic liquid.  The expression \intph\ can be useful if we want to calculate the
scattering phase in more general, time-dependent, tachyon backgrounds.
In the case of a stationary Fermi sea   the density is given by
  \eqn\statro{ \r(\xp,\xm) = \cases{1& if $ \mu< \xp\xm<\tilde\L $\
  ;\cr 0& otherwise.} }

 \subsec{Evaluation of the scattering phase  in the energy space}

\noindent 
The scattering amplitude for the eigenstates with given energy
\adjwf\ is obtained by applying the integral transformation \Mellin\
to both arguments of ${R^{ \rm ad}} (\xip,\xim) $:
\eqn\phasee{\eqalign{ \tilde R^{\rm ad}({\tilde E}_+ ,{\tilde E}_- ) &
= {1\over \pi^2} {\sinh(\pi{\tilde E}_+ )\sinh(\pi{\tilde E} _-) }
\int_0^{\infty} d\xip d\xim \xip^{i{\tilde E} _+} \xim^{i{\tilde E}
_-} \ e^{iS(\xip,\xim)} .  } }
For a stationary Fermi sea the r.h.s. depends on $\xip$ and
$\xim$ only through the variable $s$ defined in \defs, hence the
l.h.s. contains a delta function imposing the energy conservation:
\eqn\phass{\eqalign{ \tilde R^{\rm ad}({\tilde E}_+ ,{\tilde E}_- )&=
  e^{-i\d^{\rm ad}({\tilde E}_+ )  }\ \delta({\tilde E} _+-{\tilde E} _-);\cr 
e^{-i\d^{\rm ad}({\tilde E} )  }& = \frac{1}{  4\pi^2}  \,
e^{2\pi{\tilde E} } \,   \mu^{i{\tilde E} +1}\,
\int_{-\infty}^{\infty} ds\, 
e^{2 s (i{\tilde E} +1) +i {1\over \pi} (s - {1\over 2} \log{\tL \over\m})^2
- i f(-  s)}\,
.
}} 
In the last equation we replaced the sine function in front of the
integral by exponent, which is justified for large positive energies,
${\tilde E} \sim {1\over 2\pi} \log (\tL /\m)$.

We subtract, as in \Malong, the energy gap that separates the singlet
and the adjoint sectors and introduce the shifted energy\foot{Note
that our energy variable $\tilde E$ differs from the variable $\e$
used in \Malong\ by a cutoff-dependent constant, ${\tilde E} = \e +
{1\over 2\pi}\log\tL $.}
\eqn\defeh{
\hat\e = {\tilde E}  +{1\over 2\pi} \ln{\mu\over \tL } 
.} 
Then we write the integral \phass\ in the form
\eqn\adjP{\eqalign{ e^{-i\d^{\rm ad}({\tilde E})}= & \frac{1}{ 4\pi^2}
\, \tL \ \mu^{i{\tilde E} } e^{2\pi\eh} \ e^{{i\over 4\pi}( \log{\tL
\over\mu})^2} \int_{-\infty}^{\infty } ds\, e^{ 2s (i \hat \e +1) + i
f( s) - i\pi/6} , }}
where we  used the property
$
f(-x)+f(x)=  {x^2\over \pi}+{\pi\over 6}
$. 
The integral can be taken exactly using the remarkable fact (see
Appendix B) that the function $e^{ 2s + i f\(s\)}$ reproduces itself
after Fourier transformation.  Thus we obtain for the phase factor
\eqn\pasef{ \eqalign{ e^{-i\d^{\rm ad}({\tilde E})}&\sim \tL \,
\mu^{i{\tilde E}} e^{{i\over \pi}( {1\over 2\pi} \log{\tL
\over\mu})^2} e^{- i f(-\pi\eh)- i\pi/6} \cr & = \tL \, e^{ i\pi
({1\over 2\pi} \log{\tL } )^2} \times e^{ - i \pi ({\tilde E} -
{1\over 2\pi} \log{\tL } )^2 - i\pi} \times e^{ i f(\pi\eh)} .  } }
We get for the scattering phase, neglecting the large cutoff-dependent
constant,
\eqn\phaseE{\d^{\rm ad}({\tilde E})=
     \pi ({\tilde E} -  \frac{1}{ 2\pi} \log{\tL  }  )^2 
 -  f(\pi\eh).
 }
The answer is in accord with Maldacena's calculation of the scattering
phase based on the FZZ formula for the boundary two-point function in
Liouville theory, eq.  (A.4) in Appendix A of \Malong.  Comparing the
quadratic cutoff-dependent factors, we see that the cutoff $\tL$ in
the integral over energies should be identified with the square of the
large boundary cosmological constant of the FZZT brane, $\tL=
\mu_B^2$.

\newsec{Discussion}

\noindent
In this paper, we explained how to use the chiral formalism of MQM in
order to evaluate the scattering amplitudes in the non-singlet sector.
The scattering operator is given by matrix Fourier transformation and
its matrix elements can be in principle evaluated by applying recently
discovered generalizations of the Harish-Chandra-Itzykson-Zuber
formula.  We considered in detail the scattering process in which the
incoming and the outgoing states are in the adjoint representation.
The wave function in the adjoint describes a particle-like excitation
interacting with the Fermi sea, which was identified in \Malong\ as a
long folded string stretching from infinity.

We showed that the expression for the scattering amplitude such an
`adjoint particle' can be written in the form of a Fredholm
determinant.  The Fredholm kernel depends on the profile of the Fermi
sea of the singlet sector.  In the case of stationary Fermi surface,
$x^+x^-=\mu$, we reproduced Maldacena's phase shift for a single long
string \Malong.  Our tree-level formula \intph\ for the scattering
phase as an integral over the Fermi sea is actually valid also for
time-dependent perturbations of the Fermi surface, {\it e.g.}, by a
tachyon source.  This more general expression can be used to evaluate
the amplitudes of emission or absorption of tachyons by the long
string.

The scattering amplitudes involving $p$ adjoint particles are given by
the inner products of the wave functions \wfgen.  In the case when
$(\s^+)^{-1} \s^-=1$, the tree-level amplitude factorizes into $p$
one-particle amplitudes.  In the general case, when $(\s^+)^{-1} \s^-$
is a permutation with $n$ cycles, the scattering amplitude factorizes
into a product of one-cycle amplitudes.  For each such amplitude one
can use the result of \EO\ to express it as a sum of products of
one-particle amplitudes:
\eqn\Eynar{R(\x^+_1, ..., \x^+_k; \x^-_1, ..., \x^-_k) =
\sum_{\sigma\in S_k}\,
C^{(k)}_\sigma(\x_1^+,\x_1^-,\dots,\x_k^+,\x_k^-)\,\, \prod_{i=1}^k
R^{\rm ad} (\x_i^+,\x_{\sigma(i)}^-).  }
In this formula all the dependence on the eigenvalue distribution is
contained in the one-particle amplitudes, while the coefficients are
universal and related only to the $U(N)$ integration.  For example,
the one-cycle-of-length-two amplitude is given by
 \eqn\Eynrd{R(\x^+_1, \x^+_2; \x^-_1,\x^-_2) = {R^{\rm ad}
 (\x_1^+,\x_{1}^-) R^{\rm ad} (\x_2^+,\x_{2}^-) - R^{\rm ad}
 (\x_1^+,\x_{2} ^-) R^{\rm ad} (\x_2^+,\x_{1}^-) \over ( \x_1^+ -
 \x_2^+ )( \x_1^-- \x_2^-)}.  }

One can evaluate in this way the most general scattering process
involving any number of tachyons and adjoint particles.  The
corresponding asymptotic states are obtained by replacing in \wfgen\
the last factor, the ground state singlet wave function, with an
excited singlet state.
  
 An  alternative approach  to study the scattering in the non-singlet sector  
 consists in diagonalizing  the scattering matrix  given by the 
inner product  \inner.  This is possible because the problem is integrable.
    In the case of the ``upside-up'' matrix
oscillator such an approach, based on a spectrum-generating algebra
 that generalizes $W_\infty$ of the singlet sector, was developed
recently by Y. Hatsuda and Y. Matsuo \HM. In the ``upside-down" case the generators of this
algebra  create  non-singlet discrete states\foot{We
recommend \Jevicki\ as a good review paper about the role of the
$W_\infty$ symmetry and the discrete states in the singlet sector of
MQM.}  and can be used to write down Ward identities for
the $S$-matrix elements.

\bigskip
\noindent
{\bf  Acknowledgments}
\smallskip
{\ninepoint 
\noindent 
This work started as a joint project with Y. Matsuo.  I am grateful to
S. Alexandrov, V. Kazakov, N. Orantin, D. Volin and especially Y.
Matsuo for valuable comments and suggestions.  It is a pleasure to
thank Theoretical Physics Laboratory at RIKEN (Wako) and High Energy
Physics Theory Group at University of Tokyo (Hongo), where this work
was finished, for hospitality.  This research is supported by the
European Community through RTN EUCLID, contract HPRN-CT-2002-00325,
and MCRTN ENRAGE, contract MRTN-CT-2004-005616, and by the French and
Japanese governments through PAI Sakura.  }

\appendix{A}{Quasiclassical calculation of the integration kernel}

\noindent Consider first the simplest problem, the calculation of the
diagonal $E'=E$ of the kernel \defK\ for large negative energy, $-E\gg
1$.  Introduce the parametrization
\eqn\xxi{ \xpm = \sqrt{r}\, e^{ \pm\t} , \qquad \xipm =\sqrt{\r}\,
e^{\pm \s} }
and evaluate the $\t$-integral in the numerator of \bioO:
\eqn\Kmmu{\eqalign{ &\int^{\infty}_0 d \xp d \xm (\xp\xm)^{iE - 1/2}
{e^{i \xp\xm}\over (\xip+\xp)(\xim+\xm)} \cr &= \int _{-\infty}^\infty
d \t \int_{0}^\infty {d r \over \sqrt{r}}\, r^{iE }e^{i r} \, {1 \over
r+\r +2\sqrt{r\r} \cosh(\t-\s)} \cr &=\int_{0}^\infty {d r \over
\sqrt{r}}\, r^{iE }e^{i r} \, {\ln {r}-\ln{ \r} \over r-\r} .  } }
The remaining integral in
\eqn\KerN{\eqalign{ K(E,E)&= {\int_{0}^\infty {d r \over
\sqrt{r}}\, r^{iE }e^{i r} \ {\ln {r}-\ln{ \r} \over r-\r} \over
\int_{0}^\infty {d r \over \sqrt{r}}\, r^{iE }e^{i r} } } }
can be evaluated semiclassically.  Up to $o({1\over E})$ correction
the integrals in the numerator and in the denominator are saturated by
the same saddle point, $r=-E$.  Therefore in the leading order
\eqn\sdpS{ K(E,E) = {\ln(-E) -\ln \r\over -E - \r} =
  {\ln(-E) - \ln( \xi_+ \xi_-) \over -E- \xi_+ \xi_-} .  }
The off-diagonal elements $K(E, E')$ are evaluated similarly.  The
$\t$-integral in this case gives
\eqn\inttta{ \int _{-\infty}^\infty d \t {e^{2i \e\t} \over r+\r
+2\sqrt{r\r} \cosh(\t-\s)} = {\pi \over r-\r} {(r/\r)^ {i\e} - (r/\r)^
{-i\e} \over i \sinh 2\pi \e } \, e^{2i\e \s} }
and the saddle point approximation of the integral in $r$ yields
\Kernell .

 \appendix{B}{Properties of the function \deff}
 
% Some properties of this function follow from its expression in terms
% of dilogarithm:
%%
%  $$\eqalign{ f(x) &={\pi^2\over 6}+ {x^2\over\pi} + {1\over \pi}
%  \log \left(1-e^{-2 x}\right) x-{1\over2 \pi} {\Li}_2\left(e^{-2
%  x}\right)\cr &={\pi^2\over 6}+ {x^2\over\pi} + {1\over \pi}
%  -{1\over 2\pi} \int_x^\infty {y e^{-y}\over 1- e^{-y}} .} $$

 \noindent 
 1) Symmetry:
 
\eqn\fprop{ 
f(-x)+f(x)=  {x^2\over \pi}+{\pi\over 6}\, . 
}

% The function $f(x)$ can be expressed in terms of the odd function
% $g(y)$ defined as
%%
%\eqn\defg{ g(y)= - { 1\over \pi} \int_0^y dy' {y'\over \tanh y'} = -
%g(-y) .  }
%%This function is related to double sine function $S_b$ with $b=1$:
%%$${\log}S_1(1-i\a)= - {i\over 2} \int_{0}^{\infty}\frac{dt}{t}
%%\left\lbrack\frac{{\sin} [2 \a t]} {{\sinh}^2{t}) }-
%%\frac{2\a}{t}\right\rbrack = -i\pi \int_0^\a d\a' {\a'\over \tanh
%%\pi \a'} = i g ( \pi \a) $$
%The functions $f$ is related to $g$ by $$ \eqalign{ f(x)=
%%  {2\over \pi} \int _{-\infty}^x {ydy\over 1- e^{-y}} =
%{\pi\over 12} - g(x) + {x^2\over 2\pi}.  } $$ From this
%representation one finds that \eqn\fprop{ f(-x)+f(x)= {x^2\over
%\pi}+{\pi\over 6}.  }

  \bigskip

\noindent 2) Series expansion at $x\to +\infty$: \eqn\fseries{ f(x) =
{\pi^2\over 6}+ {x^2\over\pi} - \sum_{n\ge 1} \( {x\over n} + {1\over
n^2}\) e^{- 2 n x}.  }

 \bigskip
 
 \noindent 
 3)  Functional  identity:
\eqn\funcef{\eqalign{ 
\( e^{-i\pi  \p_x}   + e^{2x}  \) e^{if  (x)} &=e^{if (x)}, \cr 
%\( e^{i\pi  \p_x}   + e^{2x}  \) e^{-if  (x)} &=e^{-if (x)},  \cr 
%\( e^{i\pi  \p_x}   + e^{-2x}  \)e^{if  (-x)} &=e^{if (-x)},  \cr 
%\( e^{-i\pi  \p_x}   + e^{-2x}  \) e^{-if  (-x)} &=e^{-if (-x)}. 
}} 
 
 \bigskip
  
 \noindent 
4) Fourier transform:

\eqn\foireh{ \eqalign{ \int_{-\infty}^{\infty } d s\, e^{2s }\, e^{2is
\hat\e + i f\(s\)} =i e^{-2\pi \eh}\, e^{- if(- \pi \eh ) }.  } }

  \bigskip
  
  \bigskip

The symmetry property \fprop\ and the series expansion \fseries\
follow directly from the definition \deff.  The functional equation is
equivalent to
\eqn\liner{ f (x -i\pi ) - f(x) =- i \int ^{x}_{-\infty} d x (\coth
(x) +1) = -i \log (1-e^{2x}).  } We do not have a complete analytic
proof for the fourier transform \foireh , so we had to complete our
analytic argument with numerical evaluation.  Equation \funcef\ is symmetric under Fourier
transformation
$$
e^{if (x )} ={1\over 2\pi} \int_{\IR} dy \, e^{-{2\over\pi} i px +
i\tilde f(y)}, \quad e^{ i\tilde f(y)} = \int_{\IR} d\hat s \,
e^{{2\over\pi} i px +i f(x )} .
$$
Therefore the Fourier image satisfies the same functional equation.
There are two obvious solutions, $\tilde f(p) = f(p) $ and $\tilde
f(p) = -f(-p) $, up to a periodic function under $p\to p+i\pi$.  The
saddle point evaluation of the integral, valid for large negative $p$,
is compatible with the second choice,
$$\tilde f( p) = - f(- p) - i \ln h(p), $$
where $h$ is periodic, $h(p+i\pi)= h(p) $.
Further, the function $h(p) $ must vanish exponentially at $p\to
-\infty$, where the quasiclassics holds and we know that $ \tilde
f(p)\sim - {p^2\over \pi}$.  Therefore $h$ is expanded at $p\to-
\infty$ as a power series in $e^{2p}$.  Finally, since $f(x)$ vanishes
exponentially for large negative $x$, the function $h(p)$ must have a
simple pole ${\pi\over 2ip}$ at $p=0$.  These condition fix the form
of function $h$ as
$$
h(p) =- i {\pi} { 1 + h_1 e^{2p} +h_2 e^{4p}+\cdots \over e^{2p}-1}\ .
$$
The numerical evaluation of the integral  (we thank D. Volin for the
help with that) is compatible with
$h_1=h_2=\dots=0$.  Therefore
\eqn\dima{ h(p) = {i \over e^{2p} -1 }, } which implies
\eqn\fourief{ e^{i \tilde f(p)} := \int_{\IR} dx \, e^{{2\over\pi} i
px +i f(x )}= -i\pi \, {e^{ - i f(-p) }\over e^{2p}-1} .  } Applying
the shift equation \liner\ we get \foireh.
 
 Equation \foireh\  is actually a particular case of the formula for the Fourier transformation of the quantum dilogarithm presented in the Appendix of ref. 
 \FKV.

\listrefs \bye